# Toward a W4-F12 approach: Can explicitly correlated and orbital-based ab initio CCSD(T) limits be reconciled?

*Nitai Sylvetsky,[a] Kirk A. Peterson,[b] Amir Karton,[c] and Jan M.L. Martin*[a]*

(a) Department of Organic Chemistry, Weizmann Institute of Science, 76100 Reḥovot, Israel. Email: gershom@weizmann.ac.il. FAX: +972 8 934 3029

(b) Department of Chemistry, Washington State University, Pullman, WA 99164-4630. Email: kipeters@wsu.edu FAX: +1 509 335 8867

(c) School of Chemistry and Biochemistry, The University of Western Australia, Perth, WA 6009, Australia. Email: amir.karton@uwa.edu.au

**ABSTRACT**

In the context of high-accuracy computational thermochemistry, the valence CCSD correlation component of molecular atomization energies present the most severe basis set convergence problem, followed by the (T) component. In the present paper, we make a detailed comparison, for an expanded version of the W4-11 thermochemistry benchmark, between on the one hand orbital-based CCSD/AV{5,6}Z+d and CCSD/ACV{5,6}Z extrapolation, and on the other hand CCSD-F12b calculations with cc-pVQZ-F12 and cc-pV5Z-F12 basis sets. This latter basis set, now available for H–He, B–Ne, and Al–Ar, is shown to be very close to the basis set limit. Apparent differences (which can reach 0.35 kcal/mol for systems like $CCl_4$) between orbital-based and CCSD-F12b basis set limits disappear if basis sets with additional radial flexibility, such as ACV{5,6}Z, are used for the orbital calculation. Counterpoise calculations reveal that, while TAEs with V5Z-F12 basis sets are nearly free of BSSE, orbital calculations have significant BSSE even with AV(6+d)Z basis sets, leading to non-negligible differences between raw and counterpoise-corrected extrapolated limits. This latter problem is greatly reduced by switching to ACV{5,6}Z core-valence basis sets, or simply adding an additional zeta to just the valence orbitals. Previous reports that all-electron approaches like HEAT lead to different CCSD(T) limits than "valence limit+CV correction" approaches like FPD and W4 theory can be rationalized in terms of the greater radial flexibility of core-valence basis sets. For (T) corrections, conventional CCSD(T)/AV{Q,5}Z+d calculations are found to be superior to scaled or extrapolated CCSD(T)-F12b calculations of similar cost. For a W4-F12 protocol, we recommend obtaining the SCF and valence CCSD components from CCSD-F12b/cc-pV{Q,5}Z-F12 calculations, but the (T) component from conventional CCSD(T)/aug'-cc-pV{Q,5}Z+d calculations using Schwenke's extrapolation; post-CCSD(T), core-valence, and relativistic corrections are to be obtained as in the original W4 theory. W4-F12 is found to agree slightly better than W4 with ATcT (active thermochemical tables) data, at a substantial saving in CPU time and especially I/O overhead. A W4-F12 calculation on benzene is presented as a proof of concept.

**Introduction**

Computational thermochemistry is a cornerstone of computational chemistry, and molecular total atomization energies (TAEs, or their cognates, molecular heats of formation) are the most fundamental thermochemical properties of molecules.

A number of composite ab initio thermochemistry schemes (for reviews see Refs.[1–5]) has been developed that strive to yield such properties with 'chemical accuracy' (traditionally defined as 1 kcal/mol). These include the Gaussian-$n$ methods such as G3 and G4,[3,6] the CBS approaches of the Wesleyan U. group,[7,8] and the ccCA approach of Wilson and coworkers,[9–11] as well as Weizmann-1 (W1) theory[12–14] and its variants.[15–17] (We note here that the term 'theory', originally introduced for the 'Gn theory' family by the Pople group and followed by other groups such as that at Weizmann, is somewhat infelicitous and that terms like 'prescription', 'approach', or 'protocol' would be more appropriate: this 'theory' usage is followed here only for historical reasons.)

For more accurate calculations, there are approaches such as Weizmann-4 (W4) theory,[18,19] the HEAT approach,[20,21] and the Feller-Peterson-Dixon (FPD) approach.[22–26] The stated goal here (e.g., of W4 theory[18]) is three-sigma accuracy of 1 kJ/mol (0.24 kcal/mol) for small molecules: in the event W4 calculations would become technically feasible on medium-to-large size molecules, a modified goal of $3\sigma=0.24$ kcal/mol per (single or multiple) bond would presumably be more realistic.

In an idealized scenario, quantum chemists would be able to calculate the total atomization energy of a molecule relativistically, and including diagonal Born-Oppenheimer corrections at the FCI (full configuration interaction) or at least CCSDTQ (coupled cluster[27] with all single, double, triple, and quadruple substitutions) level near the 1-particle basis set limit, and correlating all inner-shell as well as valence electrons. In the real world, the 'scaling wall' of high-level correlated methods such as CCSDTQ — for which the CPU time requirements asymptotically scale as $O(n^4 N^6)$, n being the number of electrons and N the number of basis functions — make such calculations impossible for all but the smallest systems.

Instead, high-level composite *ab initio* methods such as W4 theory, FPD, and HEAT rely on decompositions such as:

$$TAE_e = TAE[CCSD(T)] + TAE[T_3\text{-}(T)] + TAE[T_4] + TAE[T_5] + TAE[\text{rel.}] + TAE[SO] + TAE[DBOC]$$

in which $TAE_e$ represents the total atomization energy of the molecule in the hypothetical motionless state ("at the bottom of the well") and the right-hand terms are, respectively, the all-electron CCSD(T)[28,29] atomization energy, the (usually repulsive) correction for higher-order connected triple excitations $T_3$, the (universally attractive) correction for connected quadruple excitations $T_4$, that of connected quintuple and higher excitations $T_5$, the scalar relativistic correction, the spin-orbit coupling correction, and the diagonal Born-Oppenheimer correction.

$\Delta TAE[DBOC]$ is negligible for heavy-atom systems, and quite small even for hydrogen compounds: for benzene, it reaches 0.14 kcal/mol.[30]

For a closed-shell molecule, $\Delta TAE[SO]$ is just a sum of small atomic spin-orbit splittings.

$\Delta TAE[\text{rel}]$ is quite small for 1st and 2nd row molecules, reaching about 2 kcal/mol for such ststems as $SiF_4$[31] and $SO_3$[32] but typically being a fraction of that. Electron correlation does have the effect[12,33,34] of reducing the relativistic correction by about 20%: the cross-coupling with higher-order correlation effects is negligible.

For systems dominated by a single reference determinant, the higher-order correlation terms $TAE[T_3–(T)] + TAE[T_4] + TAE[T_5]$ largely cancel, while even for molecules like ozone, their contribution is no larger than a few kcal/mol.[18,19] Their basis set convergence has been studied in some detail.[19] For the $T_3$–(T) and $T_4$ terms, basis set convergence in terms of the maximum angular momentum L was found[19] to be similar to the leading $L^{-3}$ behavior seen for the overall correlation energy.[35–37] For higher substitution levels, not only do the contributions rapidly decay but their basis set convergence becomes ever faster.[19] The $T_5$ contribution, for example, is already captured adequately by an unpolarized double-zeta basis set --- apparently these high connected excitation levels primarily reflect static rather than dynamical correlation. (See also Ref.[38] for a more detailed discussion.) This is fortunate in view of the ever-steeper computational cost scaling of these terms.

This leaves us with TAE[CCSD(T)] as the main objective, with an asymptotic $O(n^3N^4)$ cost scaling. In the HEAT approach, no further decomposition is made, and inner-shell electrons are correlated throughout. For first-row systems, this does not entail a severe computational premium, but for molecules with several second-row atoms such as $P_4$ or $AlCl_3$, the additional CPU cost and resource overhead of correlating the inner-shell electrons quickly makes the calculation intractable, especially in light of the need to use core-valence basis sets throughout. In W4 theory (and generally also FPD), a further decomposition is introduced:

$$TAE[CCSD(T)] = TAE[SCF] + TAE[CCSD, valence] + TAE[(T), valence] + TAE[CCSD(T), core-valence]$$

where the respective terms are the CCSD (coupled cluster with all singles and doubles[39]) valence correlation energy, the quasiperturbative contribution of connected triple excitations (see references for the quasi-fourth-order[40] and quasi-fifth-order[28] terms, and for the open-shell generalization[29]), and the differential inner-shell correlation contribution. The latter is dominated by the core-valence terms as the core-core correlation largely cancels between the molecule and the separated atoms.[41] Its contributions are on the order of a few kcal/mol for typical first-and second-row molecules:[12,34,42] Its basis set convergence has been studied in some detail and the conclusion was reached that extrapolation from triple- and quadruple-zeta core-valence basis sets captures the basis set limit to within a few hundredths of a kcal/mol. (We shall briefly revisit this issue in the Results and Discussion section.)

This leaves as the two largest terms the SCF and valence CCSD correlation terms, the valence (T) term typically being an order of magnitude smaller than the CCSD correlation term. Out of these, basis set convergence for the SCF term is comparatively rapid (see, e.g., Ref.[43]). Thus, the CCSD valence correlation energy can be singled out as *the* term that typically limits accuracy of ab initio thermochemical calculations. It will be the principal focus of our discussion.

Approaches such as W4 and HEAT entail extrapolation (joint in HEAT, layered in W4) to the CCSD(T) limit with basis sets as large as aug-cc-pV(6+d)Z[44–46] owing to the slow basis set convergence of the correlation energy.[35–37] In the FPD approach, basis sets as large as aug-cc-pV8Z and aug-cc-pV9Z have been used.[26,47] The requirements of these latter calculations, in terms of computation power and especially resources, make them prime candidates for convergence acceleration by means of explicitly correlated methods.[48–55] Such methods, in which "geminal" terms that explicitly depend on the interelectronic distance have been added to the orbital basis set, typically gain their users 2–3 basis set "zetas" over their conventional counterparts.[56–58]

Alas, the use of explicitly correlated methods for high-accuracy computational thermochemistry met with mixed success in the work of the present authors[15,24,59,60] and others.[61,62] While F12 methods enable rapidly reaching the vicinity of the basis set limit, approaching more closely in a consistent way proved a greater challenge. For instance, as will also be seen in this paper, basis set convergence of CCSD-F12 TAEs from F12 methods can be oscillatory or even (anomalously) monotonically decreasing, unlike the monotonically increasing behavior in conventional calculations.

Correlation consistent basis sets[44–46] have become something of a *de facto* standard for conventional *ab initio* calculations, but may not be the most suitable choice for explicitly correlated ones. In response, a team involving one of us developed the cc-pV$n$Z-F12 and cc-pCV$n$Z-F12 basis sets[63–65] ($n$ = D, T, Q), which were optimized at the MP2-F12 level in the presence of the appropriate geminal terms. These basis sets do appear to have smoother convergence behavior, though basis set extrapolation was still found to be necessary.[66]

Very recently, a cc-pV5Z-F12 basis set for the first row elements H, He, and B–Ne has been published.[67] This basis set was shown to be very close to the basis set limit, and it has proven useful in benchmarking applications involving noncovalent interactions.[67–69,70]

Application to general thermochemistry, such as a putative W4-F12 theory, would require expansion of the cc-pV5Z-F12 basis set to second row elements, as well as careful validation against a comparatively large and diverse benchmark, such as W4-11.[42]

In the process of doing so, we initially found discrepancies that appeared to suggest that CCSD and CCSD-F12b converge to different basis set limits. Upon further exploration, reported in the present paper, we found that the discrepancy was an artifact of inadequate radial flexibility of the conventional valence basis sets.

**Methods**

<u>Selection of the molecules</u>

We started with the W4-11 set.[42] We removed the three beryllium-containing compounds from the list and added the following 14 species: $C_2Cl_2$, $HC_2Cl$, $CH_3Cl$, $CH_2Cl_2$, $CHCl_3$ (chloroform), $CCl_4$ (carbon tetrachloride), FNO, ClNO, $COCl_2$ (phosgene), $CF_2Cl_2$ (Freon-12), $CHF_3$, $C_2H_3Cl$, $C_2H_5Cl$, and $CH_3CONH_2$ (formamide). For all of these, we were able to obtain CCSD-F12b results through aug-cc-pwCV5Z, as well as conventional CCSD results with various augmented basis set combinations. This extended W4-11 dataset, now totaling 151 molecules, will be denoted as W4-15 throughout.

For a subset of 28 first-row systems, denoted TAE28,[67] we previously (in the framework of Ref.[67]) obtained CCSD-F12b data with a large *sdpfgh* reference basis set (denoted REF-h),[66] as well as truncations of the same at f and g functions (REF-f and REF-g, respectively). Triple excitation contributions were obtained from conventional CCSD(T) through REF-i (i.e., adding four *i* functions to the REF-h basis set). The molecules in question are: BF, $BH_3$, BH, BN, $C_2H_2$, $C_2H_4$, $C_2$, $CF_2$, $CH_2NH$, $CH_4$, $CO_2$, CO, $F_2O$, $F_2$, $H_2CO$, $H_2O_2$, $H_2O$, $H_2$, HCN, HF, HNC, HNO, HOF, $N_2O$, $N_2$, $NH_3$, $O_3$, and $CH_2(^1A_1)$. In the present work, we are expanding TAE28 to the TAE42 set by adding in 14 second-row molecules, namely $Cl_2$, ClF, $P_2$, SiO, AlF, AlCl, CS, HCl, $H_2S$, $PH_3$, ClCN, OCS, HOCl, and $SO_2$.

Computational details

Most calculations were carried out using MOLPRO 2012.1[71] running on the Faculty of Chemistry HPC cluster at the Weizmann Institute of Science. Post-CCSD(T) correlation calculations are reported for some systems (notably benzene): these were obtained using the MRCC program of Kallay and coworkers[72] running on a cluster at the University of Western Australia.

For the cc-pV$n$Z-F12 correlation consistent basis sets ($n$ = D, T, Q) optimized for F12 calculations,[63] we employed the auxiliary basis sets[73] and CABS (complementary auxiliary basis sets)[74] developed for use with them, as well as the Weigend[75,76] JK-fitting basis sets which are the MOLPRO default. The SCF component was improved through the "CABS correction".[54,77]

For the cc-pV$n$Z-F12 basis set sequence, we have considered two ways of choosing the geminal exponents β: one the recommended[66] (rather than Ref.[63]) MP2-F12/3C(fix) optimized values of β = 0.9 for cc-pVDZ-F12,[66] 1.0 for cc-pV$x$Z-F12 ($x$ = T, Q),[66] and 1.2 for cc-pV5Z-F12;[67] the other choice being β = 1.4 throughout, as is customary with large basis sets. It appears (Table 1; see also Ref.[67] and discussion below) that using β = 1.4 throughout leads to more rapid basis set convergence at the CCSD-F12b level. It matters very little for the extrapolated V{Q,5}Z-F12 values, which are within 0.007 kcal/mol RMSD of each other for the TAE42 set, and RMSD from the reference data is 0.014-5 kcal/mol in both cases, even if the MSD does drop from 0.004 to 0.000 kcal/mol. By way of perspective: the TAE42 reference data were obtained from REF-g and REF-h basis sets,[78] the extrapolation covering just 0.015 kcal/mol RMS. For the unextrapolated V5Z-F12 basis set, however, the RMSD drops from 0.036 to 0.023 kcal/mol when β = 1.4 is chosen, while for VQZ-F12, we see a more significant lowering from 0.14 to 0.07 kcal/mol.

**Table 1. Mean signed deviations (MSD) and root mean squared deviations (RMSD) over the TAE42 dataset for CCSD-F12b valence correlation components of TAE (kcal/mol) from the best available reference calculations using large *spdfgh* basis sets.**

|  |  |  | β=1.4 | β={1.0,1.0,1.2} |  | β=1.4 |  | β=1.4 |  |  |
|---|---|---|---|---|---|---|---|---|---|---|
| MSD | F12 |  | F12 | F12 |  | F12 |  | F12 |  | orbital |
|  | MSD |  | MSD | MSD |  | MSD |  | MSD |  | MSD |
| REF-f | -0.237 | VTZ-F12 | -0.360 | -0.472 | awCVTZ | -0.524 | AV(T+d)Z | -0.353 | AV{5,6}Z separate | -0.060 |
| REF-g | -0.053 | VQZ-F12 | -0.039 | -0.116 | awCVQZ | -0.103 | AV(Q+d)Z | 0.109 | AV{5,6} Schwenke | -0.048 |
| REF-h | -0.011 | V5Z-F12rev2 | -0.009 | -0.027 | awCV5Z | -0.007 | AV(5+d)Z | 0.059 | ACV{5,6}Z | -0.016 |
| {g,h} | REF | V{Q,5}Z-F12 | 0.000 | 0.002 |  |  | AV{Q,5}Z | 0.034 | AV6Zh/ AV7Zi | -0.023 |
|  |  | V{T,Q}Z-F12 | 0.034 | 0.013 |  |  |  |  | Ditto sp[a] | -0.013 |
|  | RMSD |  | RMSD | RMSD |  | RMSD |  | RMSD |  | RMSD |
| REF-f | 0.374 | VTZ-F12 | 0.453 | 0.534 | awCVTZ | 0.616 | AV(T+d)Z | 0.506 | AV{5,6}Z separate | 0.088 |
| REF-g | 0.069 | VQZ-F12 | 0.069 | 0.136 | awCVQZ | 0.147 | AV(Q+d )Z | 0.172 | AV{5,6} Schwenke | 0.068 |
| REF-h | 0.015 | V5Z-F12rev2 | 0.023 | 0.036 | awCV5Z | 0.016 | AV(5+d)Z | 0.074 | ACV{5,6}Z | 0.045 |
| {g,h} | REF | V{Q,5}Z-F12 | 0.014 | 0.015 |  |  | AV{Q,5}Z | 0.043 | AV6Zh/ AV7Zi | 0.043 |
|  |  | V{T,Q}Z-F12 | 0.051 | 0.050 |  |  |  |  | Ditto sp[a] | 0.042 |

(a) AVn+1Z on the valence angular momenta, AVnZ for remaining angular momenta

For some systems, we also applied a large even-tempered uncontracted *spdfgh* reference basis set proposed in Ref.[78], and used in previous work by Peterson and coworkers:[79] we use the notation REF-f for its truncation at *f* functions, and similarly REF-g, and REF-h. For the REF-h basis set and its truncations, we availed ourselves of very large uncontracted auxiliary basis sets previously reported in Ref.[66]. For the V5Z-F12 basis set, we employed the combination of Weigend's aug-cc-pV5Z/JKFIT basis set[76] for the Coulomb and exchange elements with Hättig's aug-cc-pwCV5Z/MP2FIT basis set[80] for both the RI-MP2 parts and for the CABS.

For comparison, some F12 calculations were run with ordinary aug-cc-pV*n*Z basis sets,[81] where the JKFIT basis set[76] was extended by a single even-tempered layer of diffuse functions, RI-MP2 basis set was again taken from Ref.[80] but the CABS basis sets of Yousaf and Peterson[82] were employed.

The primary explicitly correlated method considered in this work is CCSD(T)-F12b,[54,55] with various forms of scaling for the connected triples. In a previous study (Ref.[67]; see also Ref.[68]), we considered CCSD(F12*) (a.k.a., CCSD-F12c)[83] instead of CCSD-F12b, and found that the difference between the two approaches is only significant for the cc-pVDZ-F12 basis set, which is manifestly inadequate for molecular atomization energies.

Basis set extrapolations for conventional calculations were carried out using Schwenke's expression[84] for CCSD and (T), while for explicitly correlated calculations we used the formulas from Ref.[78] For comparison, we also applied the original W4 scheme, in which separate $L^{-3}$ and $L^{-5}$ extrapolations are used as advocated by Klopper[46]; in this scheme, the distribution of opposite-spin, spin-up, and spin-down correlation energies to S and T pairs is not unique for open-shell systems. We followed the convention from Ref.[85]: it was found (see below) that this did not offer an advantage over Schwenke's formula, which does not require such a separation.

F12 approaches as presently practiced do not directly affect the connected quasiperturbative triples, so the basis set convergence behavior of the (T) contribution is effectively that of a conventional calculation. Marchetti and Werner[86] proposed convergence acceleration by scaling the (T) contribution by the MP2-F12/MP2 correlation energy ratio, and found that this considerably improves calculated interaction energies for noncovalent complexes. Such scaling will be indicated by the notation (T*) instead of (T). If practiced separately on molecule and separate atoms, this practice is not size-consistent: Marchetti and Werner suggested using the molecule/dimer ratio for all species, restoring size consistency — which we indicate by the suffix "sc" in (T*$_{sc}$). In two recent studies,[68,69] we found (T*) to be beneficial for F12 harmonic frequency calculations and for noncovalent interaction energies[69] as well. However, for atomization energies, Feller[87] recently performed a comparison of CCSD(T*)-F12b, extrapolated CCSD(T)-F12b and standard CCSD(T) with large basis sets for a test set of 212 molecules. He found (T*) performed well for small basis sets due to a fortuitous cancellation of errors between underestimating CCSD(corr.) and overestimating (T), a balance which disappeared with the large cc-pV5Z-F12(rev 2) basis set: Overall, CCSD(T*) provided no advantage over extrapolated CCSD(T)-F12b. In a very recent revision[70] of the S66x8 noncovalent interaction benchmark,[88] we found that the overestimate is mitigated by using the $E_{corr}$[CCSD-F12b]/$E_{corr}$[CCSD] correlation energy ratio instead, which we denote by the symbol (Tb), or (Tb$_{sc}$) for the size-consistent variant. In the original cc-pV5Z-F12 paper,[67] we instead proposed (Ts), which consists of multiplying (T) by a uniform scaling factor specific to the basis set, optimized against REF-{h,i} extrapolated values for the TAE28 set: The scaling factors thus obtained[67] are 1.1413 for VDZ-F12, 1.0527 for VTZ-F12, 1.0232 for VQZ-F12, and 1.0131 for V5Z-F12rev2. Generally, one observes ΔTAE(T) < ΔTAE(Ts) < ΔTAE(Tb$_{sc}$) < ΔTAE(T*$_{sc}$).

The MP2-F12 correlation energies discussed are those obtained with the 3C ansatz[53] with fixed amplitudes,[51] a.k.a. "3C(Fix)".

Unless noted otherwise, the "frozen core" approximation was applied, i.e., all inner-shell orbitals were constrained to be doubly occupied.

Conventional orbital-based SCF, CCSD, CCSD(T) results were obtained using the aug-cc-pV(5+d)Z[81,89] and aug-cc-pV(6+d)Z[89–91] basis sets (AV5Z and AV6Z for short), as well as with aug-cc-pCV5Z[41,92] and aug-cc-pCV6Z[64,93] core-valence basis sets (ACV5Z and ACV6Z for short) and the core-valence-weighted aug-cc-pwCVQZ and aug-cc-pwCV5Z basis sets.[41] In the conventional calculations, we omitted diffuse functions on hydrogen, a practice which has been adopted often in the past and variously denoted aug'-cc-pV$n$Z,[94] jul-cc-pV$n$Z,[95] haV$n$Z,[88] or heavy-aug-cc-pV$n$Z (e.g.,[96]). No such omission was made in the F12 calculations with ordinary aug-cc-pV$n$Z basis sets, which were carried out purely for comparison purposes.

In addition, we considered what we will denote AV$n$+1Zt basis sets, which are basis sets of the next zeta level from which the top angular momentum has been truncated. The AV5+1Zh basis set corresponds to aug-cc-pV(6+d)Z with the $i$ functions deleted (i.e., retaining at most h functions), while AV6+1Zi was derived from the aug-cc-pV(7+d)Z basis set[97–99] with the k functions removed. This constitutes an additional check of the effect of enhancing radial flexibility of the basis set (see Results and Discussion).

For benchmark conventional CCSD(T) calculations, which require no auxiliary basis sets, we also expanded the REF-h set with four additional $i$ functions[66] to obtain the REF-i basis set.

Basis set optimization

The development of the new cc-pV5Z-F12 orbital basis sets for Al–Ar was similar to the previous optimizations of the $n$ = D–Q sets.[63] In the present work the s and p portions of the basis sets were simply taken from the standard contracted aug-cc-pV6Z basis sets.[90] Higher angular momentum correlating functions optimized for the MP2-F12/3C(D) total energy[53] were then added to these HF sets, i.e., (6d4f3g2h). In each case the exponents were constrained to follow an even-tempered sequence, except for the tightest d and f functions, which were freely optimized as in the original cc-pV$n$Z-F12 optimizations.[63] In contrast however to the cc-pV5Z-F12 sets for B–Ne, the correlating functions of this work were optimized for the ground states of the atoms instead of the homonuclear diatomics. This was only a matter of convenience and was not expected to affect the quality of the resulting basis sets, particularly for one of this size. For consistency with Ref.[63], all optimizations employed a geminal exponent of 1.4 with the reference DF and RI basis sets of Ref.[66]. (This choice of the geminal exponent keeps the optimized orbital exponents somewhat more diffuse so that the F12 factor covers the short range correlation, leaving the basis set to take care of the long range.)

The geminal exponent β was optimized at the MP2-F12 level for the new cc-pV5Z-F12 sets according to the same procedure as in Ref. [66] and was found to be β = 1.2.

**Results and discussion**

Calibration against TAE42

In Table 1, we present error statistics for the TAE42 dataset, compared to basis set limits extrapolated from large, uncontracted *spdfg* and *spdfgh* basis sets proposed in Ref.[78] (denoted REF-g and REF-h for short). The basis set limit was obtained by extrapolation using the theoretical $L_{max}^{-7}$ dependence derived by Kutzelnigg.[37] As it bridges just 0.015 kcal/mol RMS, it is deemed adequate for our purposes. A conservative estimate for the uncertainty on the REF-{g.h} limits would be about 0.01 kcal/mol RMS

Extrapolation from V{Q,5}Z-F12 yields results of nearly the same quality as REF-h, RMSD = 0.014 kcal/mol with β = 1.4 , or 0.015 kcal/mol with 'optimal' geminal exponents. The raw VTZ-F12, VQZ-F12, and V5Z-F12 results are unquestionably closer to the basis set limit with β = 1.4 than with the MP2-F12 optimized β. Previous attempts to optimize geminal exponents at the CCSD-F12b level led to unrealistically high β values (see, e.g., p.8 of Ref.[78]). We showed previously[67] for REF-h and for cc-pV5Z-F12 that with sufficiently large basis sets, the dependence of the correlation energy on the geminal exponent is weak enough that optimization of β becomes pretty much irrelevant: for CCSD-F12b or CCSD(F12*) calculations in cc-pVTZ-F12 or especially cc-pVQZ-F12 basis sets — where β = 1.4 cuts the error in half, from RMSD = 0.136 to 0.069 kcal/mol — setting β = 1.4 may be a sensible choice.

CCSD-F12b/awCV5Z yields results of nearly the same quality as REF-h. However, unlike the V$n$Z-F12 series in which basis set convergence is monotonic, CCSD-F12b/awCVQZ TAEs can be larger than their CCSD-F12b/awCV5Z counterparts.

Using CCSD-F12b with conventional AV(n+d)Z basis sets yields not only non-monotonic convergence — AV(Q+d)Z and AV(5+d)Z actually overbind, on average — but even the costly AV(5+d)Z basis set still leaves an RMSD = 0.074 kcal/mol. AV{Q,5}Z extrapolation does cut this figure almost in half, but V{Q,5}Z-F12 clearly outperforms it, at comparable or lower computational cost.

Turning now to conventional CCSD calculations, the standard W4 extrapolation procedure employed in the W4-11 paper yields RMSD = 0.08 kcal/mol, with a clear underestimate on average (MSD = –0.05 kcal/mol). Switching to joint extrapolation using Schwenke's formula (and hence eliminating the ambiguity as to how to partition the open-shell correlation energies between S and T pairs) actually somewhat reduces RMSD to 0.06 kcal/mol.

Substituting ACV{5,6}Z core-valence basis sets, however, reduces both systematic and RMSD error, the latter to 0.045 kcal/mol. Using AV$n$+1Z basis sets from which the top angular momentum has been removed (denoted AV6Zh and AV7Zi in the table) actually yields a statistically equivalent RMSD of 0.043 kcal/mol. This suggests that the issue is related to greater radial flexibility in these basis sets. In order to verify whether this results primarily from more flexible valence orbitals, or also from the availability of additional polarization functions, we carried out an additional set of calculations in which AVn+1Z basis set for s and p orbitals (for H, just s orbitals) was combined with the ordinary AVnZ basis set for the remaining angular momenta. This yields essentially the same performance, confirming that insufficient radial flexibility in the valence angular momenta of the AVnZ basis sets is the primary culprit.

For V{Q,5}Z-F12, we obtain essentially the same results whether we use $\beta = 1.4$ throughout or the recommended[66] geminal exponent sequence: 1.0, 1.0, 1.2 for n = T, Q, 5, respectively. For the smaller basis sets, $\beta = 1.4$ greatly reduces the systematic error. In addition, especially for VQZ-F12, $\beta = 1.4$ causes a fairly dramatic improvement in relative terms (from RMSD = 0.134 to 0.057 kcal/mol).

V{T,Q}Z-F12 extrapolation can achieve RMSD = 0.05 kcal/mol with either extrapolation sequence, but the systematic error is somewhat smaller with the "optimal" sequence.

<u>The complete W4-15 set</u>

Let us now turn to larger datasets, namely the 137-member W4-11 and its expanded version W4-15 containing 151 molecules with up to five non-hydrogen atoms. Here, we use cc-

pV{Q,5}Z-F12 extrapolated data with β = 1.4 as a secondary standard. RMS deviations are presented in Table 2.

**Table 2. MSD and RMSD for CCSD-F12b valence correlation components of TAE (kcal/mol) if the W4-11 and W4-15 datasets. The V{Q,5}Z-F12 extrapolated data with β = 1.4 were used as the reference.**

| | F12, β=1.4 | | | | | |
|---|---|---|---|---|---|---|
| | AV{Q,5}Z | V{T,Q}Z-F12 | V{Q,5}Z-F12 | V5Z-F12 | awCV5Z | AV5Z |
| RMSD W4-11 | 0.063 | 0.060 | 0.000 | 0.024 | 0.018 | 0.107 |
| MSD W4-11 | 0.049 | 0.045 | 0.000 | -0.014 | -0.007 | 0.086 |
| RMSD W4-15 | 0.070 | 0.058 | 0.000 | 0.024 | 0.017 | 0.111 |
| | F12, β=1.0, 1.0, 1.2 | | | | | |
| | AV{Q,5}Z | V{T,Q}Z-F12 | V{Q,5}Z-F12 | V5Z-F12 | awCV5Z | AV5Z |
| RMSD W4-11 | 0.064 | 0.059 | 0.000 | 0.044 | 0.019 | 0.100 |
| MSD W4-11 | 0.046 | 0.019 | 0.000 | -0.037 | -0.010 | 0.081 |
| RMSD W4-15 | 0.071 | 0.058 | 0.000 | 0.044 | 0.019 | 0.105 |
| Orbital-only calculation | conv., split $L^{-3}$ and $L^{-5}$ (a) AV{5,6}Z | conventional, Schwenke AV{5,6}Z | joint Schwenke extrapolation | | | |
| | | | ACV{5,6}Z | AV6Zh/AV7Zi | Ditto sp[b] | |
| RMSD W4-11 | 0.101 | 0.084 | 0.056 | 0.050 | 0.051 | |
| MSD W4-11 | -0.056 | -0.054 | -0.019 | -0.027 | -0.012 | |
| RMSD W4-15 | | 0.097 | 0.055 | 0.051 | 0.050 | |

(a) Separate $L^{-3}$ extrapolation for singlet-coupled pairs, $L^{-5}$ for triplet-coupled pairs;

(b) AVn+1Z on valence occupied angular momenta, AVnZ on remainder.

For the entire W4-11 set, we find an RMSD between orbital-based AV{5,6}Z values and V{Q,5}Z-F12 of 0.10 kcal/mol, on average systematically underestimated by 0.06 kcal/mol. The RMSD can in fact be reduced to 0.08 kcal/mol by employing Schwenke's joint extrapolation, which removes the ambiguity over the S- and T-pair distribution. However, the systematic underestimate remains, and in fact the RMSD goes up again to 0.10 kcal/mol if the additional species are included. Differences are especially large for polychlorides: one particularly instructive example is CCl$_4$ (Table 3), where AV{5,6}Z differs by -0.36 kcal/mol from our V{Q,5}Z-F12 limit, which is within 0.03 kcal/mol from awCV5Z and 0.01 kcal/mol from the ACV{5,6}Z extrapolation. One sees the same to a lesser extent for polyfluorides. Switching to core-valence basis sets for the extrapolation cuts the RMSD to 0.056 kcal/mol even for the expanded W4-15 set, and the systematic bias to just -0.023 kcal/mol. Results of a similar quality can be obtained with truncated AV6Zh/AV7Zi basis sets, or indeed with AVnZ basis sets in which just the valence angular momenta were replaced by their AVn+1Z counterparts.

**Table 3. Deviations (kcal/mol) for CCl$_4$ with different basis set sequences for conventional and explicitly correlated calculations.**

| F12 AV{Q,5}Z | F12 V{Q,5}Z-F12 | F12 V5Z-F12 | F12 awCV5Z | F12 AV5Z | Orbital, joint Schwenke extrapolation | | | |
|---|---|---|---|---|---|---|---|---|
| | | | | | AV{5,6}Z | ACV{5,6}Z | AV6Zh/ AV7Zi | Ditto sp |
| 0.215 | REF | -0.034 | 0.029 | 0.180 | -0.356 | 0.005 | 0.100 | 0.000 |

What about CCSD-F12b/awCV5Z? The RMSD with V{Q,5}Z-F12 is just under 0.02 kcal/mol, buttressing the case for our V{Q,5}Z-F12 reference values. Average difference is just -0.01 kcal/mol.

Had we used unextrapolated V5Z-F12 results as is, that would have led to an RMSD = 0.04 kcal/mol and an average underestimate by almost the same amount.

In contrast, using the original AV5Z basis set in conjunction with CCSD-F12b yields an RMSD of about 0.1 kcal/mol, with a systematic overestimate by almost that amount. AV{Q,5}Z extrapolation reduces this to 0.07 kcal/mol, with still an average overestimate.

Finally, we note in passing that the use of the V5Z-F12 basis sets, without the additional polarization functions on hydrogen that are included in V5Z-F12rev2, incurs an RMSD of 0.019 kcal/mol. The additional basis functions on hydrogen cause changes as large as 0.08 kcal/mol for propane. As in the original V5Z-F12 paper,[67] we argue that the additional basis functions are strongly recommended for high-accuracy thermochemical work, even as they were found to be surplus to the requirements for noncovalent interactions[67] and to mainly cause near-linear dependence issues there.

Basis set superposition error in atomization energies

In an attempt to rationalize the above findings, Table 4 presents calculated counterpoise corrections for the dissociation energies of $N_2$, $F_2$, $P_2$, $S_2$, $Cl_2$, and $CO_2$ with various basis set sequences. For the $CO_2$ triatomic, the site-site function counterpoise method of Wells and Wilson[100] was employed, i.e., the counterpoise-corrected atomization energy of $CO_2$ was taken as $2E[O(C)(O)]+E[(O)C(O)]-E[CO_2]$, where parentheses indicate ghost atoms.

A number of observations can be made here. First of all, in non-extrapolated orbital-based calculations, the counterpoise corrections are chemically nontrivial even with basis sets as large as AV(6+d)Z.

**Table 4: CCSD level basis set superposition errors (kcal/mol) for five diatomic molecules and $CO_2$ using different basis sets, both explicitly correlated and conventional.**

| | $N_2$ | $F_2$ | $P_2$ | $S_2$ | $Cl_2$ | $CO_2$ |
|---|---|---|---|---|---|---|
| BSSE on CCSD-F12b correlation contribution to $D_e$ (kcal/mol) | | | | | | |
| β=1.4 | $N_2$ | $F_2$ | $P_2$ | $S_2$ | $Cl_2$ | $CO_2$ |
| VDZ-F12 | -0.752 | -0.414 | -1.091 | -2.462 | -1.451 | -1.729 |
| VTZ-F12 | -0.144 | -0.149 | -0.114 | -0.567 | -0.473 | -0.393 |
| VQZ-F12 | -0.027 | -0.032 | -0.023 | -0.108 | -0.092 | -0.092 |
| V5Z-F12 | -0.012 | -0.012 | 0.002 | -0.011 | -0.017 | -0.027 |
| AVDZ | -0.786 | -0.321 | -1.086 | -2.311 | -1.282 | -0.518 |
| AVTZ | -0.324 | -0.244 | -0.145 | -0.553 | -0.481 | -0.400 |
| AVQZ | -0.186 | -0.169 | -0.049 | -0.140 | -0.139 | -0.347 |
| AV5Z | -0.063 | -0.043 | -0.031 | -0.063 | -0.072 | -0.120 |
| AV{T,Q}Z | -0.128 | +0.138 | -0.009 | +0.032 | +0.003 | -0.325 |
| AV{Q,5}Z | -0.001 | +0.020 | -0.023 | -0.024 | -0.039 | -0.006 |
| BSSE on CCSD correlation contribution to $D_e$ (kcal/mol) | | | | | | |
| orbital-only | $N_2$ | $F_2$ | $P_2$ | $S_2$ | $Cl_2$ | $CO_2$ |
| AV(T+d)Z | -1.559 | -1.213 | -0.663 | -1.682 | -1.434 | -3.905 |
| AV(Q+d)Z | -0.662 | -0.657 | -0.273 | -0.653 | -0.563 | -1.749 |
| AV(5+d)Z | -0.289 | -0.282 | -0.152 | -0.359 | -0.360 | -0.794 |
| AV(6+d)Z | -0.153 | -0.150 | -0.082 | -0.192 | -0.164 | -0.427 |
| extrapolated BSSE should be as close to zero as possible | | | | | | |
| AV{Q,5}Z+d | 0.058 | 0.067 | -0.039 | -0.085 | -0.171 | 0.094 |
| AV{5,6}Z+d | 0.019 | 0.017 | 0.007 | 0.019 | 0.084 | 0.037 |
| orbital-only | $N_2$ | $F_2$ | $P_2$ | $S_2$ | $Cl_2$ | $CO_2$ |
| ACVTZ | -1.169 | -0.891 | -0.583 | -1.580 | -1.347 | -3.069 |
| ACVQZ | -0.489 | -0.467 | -0.246 | -0.617 | -0.517 | -1.341 |
| ACV5Z | -0.224 | -0.210 | -0.137 | -0.208 | -0.196 | -0.619 |
| ACV6Z | -0.122 | -0.114 | -0.078 | -0.118 | -0.114 | -0.340 |
| ACV{Q,5}Z | 0.023 | 0.029 | -0.036 | 0.172 | 0.103 | 0.053 |
| ACV{5,6}Z | 0.007 | 0.007 | 0.003 | -0.004 | -0.010 | 0.013 |
| orbital-only | $N_2$ | $F_2$ | $P_2$ | $S_2$ | $Cl_2$ | $CO_2$ |
| AVQZf | -0.888 | -0.827 | -0.515 | -1.631 | -1.469 | -2.470 |
| AV5Zg | -0.376 | -0.328 | -0.225 | -0.570 | -0.524 | -1.004 |
| AV6Zh | -0.195 | -0.165 | -0.113 | -0.277 | -0.236 | -0.516 |
| AV7Zi | -0.114 | -0.097 | -0.069 | -0.166 | -0.158 | -0.304 |
| {Q,5} | -0.027 | 0.013 | -0.009 | -0.004 | 0.032 | -0.061 |
| {5,6} | -0.011 | -0.011 | -0.013 | -0.026 | -0.059 | -0.037 |
| orbital-only | $N_2$ | $F_2$ | $P_2$ | $S_2$ | $Cl_2$ | $CO_2$ |
| AVQZf_SP | -0.979 | -0.799 | -0.551 | -1.520 | -1.329 | -2.544 |
| AV5Zg_SP | -0.427 | -0.387 | -0.224 | -0.575 | -0.523 | -1.163 |
| AV6Zh_SP | -0.239 | -0.215 | -0.111 | -0.266 | -0.227 | -0.657 |
| AV7Zi_SP | -0.137 | -0.126 | -0.069 | -0.164 | -0.142 | -0.379 |
| {Q,5} | -0.064 | -0.055 | -0.006 | 0.023 | 0.049 | -0.187 |
| {5,6} | -0.008 | -0.013 | -0.016 | -0.036 | -0.035 | -0.027 |

A positive CP contribution at the basis set limit means the CP-corrected limit is more binding than the CP-uncorrected one.

We note that in the F12 calculations, the AVnZ results carry very nontrivial HF+CABS BSSEs (unlike VnZ-F12)

Second, while in principle extrapolation to the complete basis set limit should lead to a vanishing counterpoise correction, this is manifestly not the case for the second-row species (especially $Cl_2$) with the AVnZ sequence.

Third, in the explicitly correlated calculations, cc-pV5Z-F12, in contrast, does have essentially negligible CP corrections, much unlike AV5Z when used in that context. We add that, while the BSSEs for the SCF (i.e., HF+CABS) components in the cc-pVnZ-F12 series quickly taper off to essentially zero, this is emphatically not the case for the AVnZ series.

Fourth, coming back to conventional calculations, the alternative basis set sequences ACVnZ and AV6Zh/AV7Zi, i.e., the AVn+1Z basis sets with the top angular momentum deleted, suffer noticeably less from the issue. What these two sequences have in common, for the purposes of a valence calculation, is enhanced radial flexibility. Much of the benefit is recovered, as is seen at the bottom of Table 4, by simply using AVn+1Z for the valence angular momenta in conjunction with AVnZ for the remainder — i.e., by adding a zeta to the valence orbitals.

<u>SCF component and core-valence separation</u>

SCF limits for the W4-15 dataset were established by Karton-Martin extrapolation[43] from ACV5Z and ACV6Z results. RMS deviations for various basis sets in conventional and explicitly correlated (HF+CABS) calculations are summarized in Table 5.

As can be seen there, the HF components for the ACVnZ series converge quite rapidly, with ACV5Z already within 0.01 kcal/mol RMS and the extrapolation accounting for post-ACV6Z expansion amounting to just 0.002 kcal/mol RMS. HF+CABS/VQZ-F12 is already quite close at RMSD=0.016 kcal/mol, while HF+CABS/V5Z-F12 is essentially converged with respect to the orbital basis set at RMSD=0.003 kcal/mol.

**Table 5: RMS deviations (kcal/mol) over the W4-15 set for the SCF component of the total atomization energies in the W4-15 set.**

| HF | | HF | | HF | | HF+CABS | | HF+CABS | |
|---|---|---|---|---|---|---|---|---|---|
| AVQZ | 0.264 | ACVQZ | 0.024 | | | VTZ-F12 | 0.083 | AVTZ | 0.149 |
| AV5Z | 0.098 | ACV5Z | 0.010 | AV6Zh | 0.130 | VQZ-F12 | 0.016 | AVQZ | 0.050 |
| AV6Z | 0.023 | ACV6Z | 0.002 | AV7Zi | 0.025 | V5Z-F12 | 0.003 | AV5Z | 0.017 |
| AV{Q,5}Z | 0.061 | | | | | | | | |
| AV{5,6}Z | 0.008 | ACV{5,6}Z | REFERENCE | | | | | | |

In contrast, HF/AV(5+d)Z still has an RMSD of 0.10 kcal/mol (individual errors reaching 0.6 kcal/mol for $SO_3$), and even for HF/AV(6+d)Z an RMSD = 0.021 kcal/mol remains. There has been some discussion between the W4 and HEAT groups as to the reasons for the difference between joint all-electron CCSD(T) extrapolation using core-valence basis sets (as practiced in HEAT) and the layered extrapolation of SCF, CCSD, and (T) using valence basis sets, augmented with core-valence corrections. This difference can be decomposed into three terms:

TAE[CCSD(T,all)/ACVnZ]–TAE[CCSD(T,val)/AVnZ] = TAE[CCSD(T,all)/ACVnZ]
–TAE[CCSD(T,val)/ACVnZ]+TAE[CCSD(T,val)/ACVnZ]–TAE[CCSD(T,val)/AVnZ]

=TAE[CV]+{TAE$_{corr}$[CCSD(T,val)/ACVnZ]–TAE$_{corr}$[CCSD(T,val)/ACVnZ]}

+{TAE [HF/ACVnZ]- TAE$_{corr}$[HF/AVnZ]}

where the first term, TAE[CV], represents the core-valence correction proper (calculated separately in W4 theory), the second terms is the change in the valence correlation terms due to the more flexible basis set, and the third term the effect of this basis sets expansion on the Hartree-Fock component. Based on the present results in Tables 2 and 5, we are prepared to say that the difference is primarily due to the improved valence correlation term resulting from additional radial flexibility, with the additional HF relaxation energy a secondary factor.. We note that these latter two terms are included implicitly in ccCA[11] through the technique of taking the CV correction as the difference between a core-valence calculation in a CV basis set and a valence calculation in a valence set.

As seen in the previous subsection, the SCF component is not the whole story. Detailed comparison between $TAE_{corr,val}$[CCSD] for AV(6+d)Z and ACV6Z basis sets reveals differences reaching up to 0.1 kcal/mol, with the core-valence basis set yielding less binding, owing to reduced basis set superposition error. Between AV(5+d)Z and ACV5Z, the differences are much larger, reaching 0.3 kcal/mol for $CCl_4$ and 0.24 kcal/mol for $CF_4$, again, due to reduced BSSE. The Schwenke-style extrapolation is incapable of reducing the AV{n-1,n}Z BSSE to the desired level: 0.09 kcal/mol remains for $CF_4$, 0.25 kcal/mol for $AlCl_3$, 0.15 kcal/mol for $Cl_2O$, 0.12 kcal/mol for $Cl_2$, 0.23 kcal/mol for $CF_2Cl_2$, and a whopping 0.39 kcal/mol for $CCl_4$. We note that in all these cases, the V{Q,5}Z-F12 limit or the raw V5Z-F12 results are closer to the ACV{5,6}Z answer than to the AV{5,6}Z+d one.

Finally, the question remains how well the core-valence correlation contributions themselves can be captured.

For a subset of 63 molecules, we calculated core-valence correlation contributions at the CCSD(T)/ACVnZ level (n=T,Q,5,6). The ACV{5,6}Z extrapolated values were used as a primary standard. RMS deviations using smaller basis sets are given in Table 6.

**Table 6: RMS deviations (kcal/mol) for the inner-shell correlation contribution to the total atomization energies.**

|            | 63-system subset | W4-15     |
|------------|------------------|-----------|
| aCVTZ      | 0.162            | —         |
| aCVQZ      | 0.069            | —         |
| aCV5Z      | 0.031            | —         |
| aCV6Z      | 0.016            | —         |
| aCV{T,Q}Z  | 0.041            | —         |
| aCV{Q,5}Z  | 0.014            | —         |
| aCV{5,6}Z  | REFERENCE        | —         |
| awCVTZ     | 0.108            | 0.166     |
| awCVQZ     | 0.039            | 0.066     |
| awCV5Z     | 0.018            | 0.034     |
| awCV{T,Q}Z | 0.020            | 0.021     |
| awCV{Q,5}Z | 0.009            | REFERENCE |

As can be seen above, the CCSD(T)/aug'-cc-pwCV{T,Q}Z level used for that contribution in W4 theory captures the basis set limits to within 0.02 kcal/mol RMS. The statistics are actually somewhat worsened by especially poor performance for some Al and Si hydrides: for $Si_2H_6$, for instance, the awCV{T,Q}Z and awCV{Q,5}Z limits differ by 0.10 kcal/mol. Further inspection revealed that the extrapolations were skewed by especially poor performance of the awCVTZ basis sets: the ACV{T,Q}Z sets did not have the problem. Further analysis of the basis sets revealed that the inner-shell $f$ correlation functions in awCVTZ has such a small exponent (the better to describe core-valence correlation) that it leaves the basis set insufficient to describe core-core correlation.

At any rate, awCV{Q,5}Z calculations should put that issue to rest in the event that this level of accuracy is required.

The effect of higher-order inner-shell correlation was briefly touched upon in Ref.[19]. It is effectively nil for systems dominated by dynamical correlation but can reach the 0.1-0.2 kcal/mol regime for molecules with pathological nondynamical correlation such as singlet $C_2$ and BN. They are considered in W4.4 theory, at great computational expense that effectively puts applications beyond *very* small systems out of reach.

Connected triple excitations, (T) corrections

In contrast, as summarized in Table 6, the corresponding differences for the (T) term are small, and essentially vanish upon extrapolation, both {5,6} and the smaller {Q,5}.

In other words, AV{5,6}Z, ACV{5,6}Z, and the AV6Z(no i)/AV7Z(no k) sequence all yield essentially the same values. For the TAE42 set, all three of these agree to within about 0.01 kcal/mol RMS, as do AV{Q,5}Z and ACV{Q,5}Z with Schwenke's extrapolation (or Ranasinghe and Petersson's,[101] which yields nearly equivalent results). Extrapolation in the REF-{h,i} results bridges 0.025 kcal/mol RMS.

Concerning F12 calculations, unscaled (T) is clearly unacceptable even with the V5Z-F12 basis set, at RMSD=0.18 kcal/mol for W4-15. Uniform scaling (Ts), as proposed in Ref.[67], yields RMSD=0.10 kcal/mol for VQZ-F12, and still RMSD=0.04 kcal/mol for V5Z-F12. The $(Tb_{sc})$ procedure, proposed in Ref.[70] and found to be successful there for noncovalent interactions and smaller basis sets, turns out to work less well than (Ts), at RMSD=0.06 kcal/mol; since generally (Ts) < $(Tb_{sc})$ < (T*), it does not surprise that (T*) would be even less effective.

In fact, the AV{Q,5}Z extrapolation that went into the W4-11 paper has a smaller RMSD. Simply replacing the extrapolation there by Schwenke's reduces its error to 0.008 kcal/mol RMS. Similar results are obtained for ACV{Q,5}Z.

**Table 7: RMSD (kcal/mol) from Schwenke-extrapolated AV{5,6}Z+d connected triple excitations contributions to the TAE, for both the W4-11 and W4-15 datasets**

| TAE42 | W4-11 | W4-15 | Extrapolation type | basis sets |
|---|---|---|---|---|
| 0.041 | 0.040 | 0.042 | $L^{-3}$ | AV{Q,5}Z+d |
| 0.011 | 0.008 | 0.009 | Schwenke | AV{Q,5}Z+d |
| 0.010 | 0.007 | 0.007 | Schwenke | ACV{Q,5}Z |
| 0.011 | REF | REF | Schwenke | AV{5,6}Z+d |
| 0.012 | 0.002 | 0.003 | Schwenke | ACV{5,6}Z |
| 0.009 | 0.005 | 0.005 | Schwenke | AV6Zh/AV7Zi |
| 0.089 | 0.101 | 0.107 | Ref.[70] | (T$_{bsc}$) VQZ-F12 |
| 0.062 | 0.063 | 0.067 | Ref.[70] | (T$_{bsc}$) V5Z-F12 |
| 0.066 | 0.096 | 0.097 | ×1.0232[67] | (Ts) VQZ-F12 |
| 0.031 | 0.042 | 0.042 | ×1.0131[67] | (Ts) V5Z-F12 |
| 0.133 | 0.173 | 0.176 | none | (T) V5Z-F12 |
| 0.037 | 0.047 | 0.045 | Peterson | V{T,Q}Z-F12 |
| 0.045 | 0.039 | 0.038 | Peterson-like | V{Q,5}Z-F12 |
| REF | N/A | N/A | $L^{-3}$ | REF-{h,i} |
| 0.025 | N/A | N/A | *none* | REF-i raw |
| 0.044 | N/A | N/A | *none* | REF-h raw |

We conclude that for accurate thermochemical work, it is best to obtain (T) separately from orbital calculations, preferably AV{Q,5}Z+d or better, and add that on to CCSD-F12b/cc-pV{Q,5}Z-F12 results.

Application to larger systems: benzene

A full W4 calculation on benzene is precluded by near-singularity issues in the aug-cc-pV6Z basis set for benzene, as was also seen in recent studies by Harding et al. (HVGSK)[30] on benzene itself, and by Xantheas[102] on benzene dimer. This is obviously to some extent a result of the lack of need for diffuse functions with this large basis set, particularly for H. We were able to carry out CCSD/cc-pV6Z calculations without any issues though. Our data below are at the CCSD(T)/cc-pVQZ geometry, $r_{CH}$ = 1.08260 Å and $r_{CC}$ = 1.39498 Å, which differs from the reference geometry of Harding et al. in that the C(1s) core electrons were frozen in the geometry optimization.

The RHF/cc-pV6Z atomization energy, 1044.97 kcal/mol, agrees very closely with the HF+CABS/cc-pVQZ-F12 value of 1044.99 kcal/mol, and its cc-pV5Z-F12 counterpart of 1045.01 kcal/mol.

At the CCSD/cc-pV5Z and CCSD/cc-pV6Z levels, we obtain valence correlation contributions of 286.04 and 288.14 kcal/mol, respectively, leading to an extrapolated CCSD/cc-pV{5,6}Z basis set limit of 290.72 kcal/mol using Schwenke's extrapolation.

At the CCSD-F12b/cc-pVQZ-F12 level, we obtain 290.594 kcal/mol; with the cc-pV5Z-F12 basis set, this rises to 290.683 kcal/mol, leading to a CCSD-F12b/cc-pV{Q,5}Z-F12 limit of 290.71 kcal/mol. The orbital-based and F12 calculations are thus seen to be in perfect agreement.

The calculated (T)/cc-pV$n$Z ($n$ = Q, 5) energies of 25.908 and 26.362 kcal/mol translate into a Schwenke extrapolated value of 26.70 kcal/mol. The unscaled (T) contribution from CCSD(T)-F12b/cc-pV5Z-F12 is 26.294 kcal/mol. Marchetti-Werner scaling[86] (based on the molecular $E_{corr}$[MP2-F12]/$E_{corr}$[MP2] ratio) leads to (T*$_{sc}$) = 26.978 kcal/mol; in contrast, (Tb$_{sc}$) obtained from the $E_{corr}$[CCSD-F12b]/$E_{corr}$[CCSD], as proposed in Ref.[70] clocks in at ($Tb_{sc}$) = 26.828 kcal/mol, much closer to the conventional value. Uniform scaling of the triples, as proposed by us in Ref.[67] leads to a smaller (Ts) = 26.651 kcal/mol.

The valence CCSD(T) limit works out to 1362.42 kcal/mol combining CCSD-F12b/cc-pV{Q,5}Z-F12 with (T)/V{Q,5}Z. Core-valence correlation at the CCSD(T) level adds 6.670 (awCVTZ) and 7.074 (awCVQZ) kcal/mol, which extrapolates to 7.369 kcal/mol at the basis set limit. (The Schwenke and $L^{-3}$ extrapolations are essentially equivalent for that basis set pair.) This works out to a CCSD(T) all-electron limit of 1369.79 kcal/mol, about 0.5 kcal/mol lower than the HVGSK limit of 1370.3 kcal/mol (Table IV in that work). After adding in post-CCSD(T) correlation contributions, scalar relativistic effects, atomic spin-orbit splitting, diagonal Born-Oppenheimer corrections, and the anharmonic zero-point vibrational energy, those authors end up with a TAE of 5463.0±3.1 kJ/mol, i.e., 1305.7±0.74 kcal/mol, which is in excellent agreement with the latest ATcT (Active Thermochemical Tables[103–109]) [[http://atct.anl.gov ver. 1.112]] value of 1305.9±0.1 kcal/mol.

The contribution of fully iterative triples was found as -2.62 kcal/mol in this work, extrapolated from cc-pV(D,T}Z. This is not greatly different from -2.68 kcal/mol obtained by HVGSK. Those authors treated $T_4$ as a CCSDT(Q)/cc-pVDZ value, essentially identical to our recalculation at the present geometry of +1.63 kcal/mol.

We were, however, able to complete a CCSDT(Q)/cc-pVTZ calculation on benzene, which entailed 2.2 *trillion* (Q) contributions and 3.1 billion iterative CCSDT amplitudes. This led to 1.9823 kcal/mol. As we have previously shown[19] that the basis set convergence of (Q) empirically follows a similar $L^{-3}$ pattern as the CCSD correlation energy, the extrapolated cc-

pV{D,T}Z value becomes +2.13 kcal/mol. Thus, the gap between the respective CCSD(T) limits is almost exactly compensated by the gap in the post-CCSD(T) corrections.

At the end of the day, both groups are in excellent agreement with the ATcT determination.

**A W4-F12 PROTOCOL AND ITS PERFORMANCE**

The W4-F12 protocol is now defined as follows:

• CCSD contribution at the CCSD-F12b/cc-pV{Q,5}Z-F12 level with $\beta = 1.4$

• SCF contribution from HF+CABS with the cc-pV5Z-F12 basis set in the CCSD step

• (T) extrapolated using Schwenke's formula from CCSD(T)/AV{Q,5}Z+d basis sets

• all remaining steps as in the original W4 protocol

In Ref.[42], for the ATcT data available at the time (only species with ATcT uncertainty 0.10 kcal/mol or less were considered), the RMSD for W4 theory was found to be 0.102 kcal/mol. In the present work, we apply inverse-variance weighting to the experimental data, except that we apply a lower bound of 0.005 kcal/mol to the uncertainties, lest the very precise experimental values for a few diatomics overwhelm the other data. Thus, we obtain a weighted RMSD of 0.105 kcal/mol for straight W4, compared to 0.071 kcal/mol for W4-F12, 0.085 for W4 with ACVnZ basis sets, and 0.080 kcal/mol for W4 with one extra zeta in the sp parts of the basis sets. Adjusting the lower bound downward increases the gap in favor of W4-F12: for instance, with a lower bound of 0.002 kcal/mol we obtainRMSD=0.055 kcal/mol for W4-F12, compared to 0.099 kcal/mol for straight W4 and 0.064 kcal/mol for W4 with an extra zeta in sp. If instead we adjust the lower uncertainty bound upward to 0.01 kcal/mol, W4, W4-F12, and W4 with an extra sp zeta clock in at 0.11, 0.085, and 0.079 kcal/mol, respectively.

Particularly satisfying is the reduction in the error for the accurately known dissociation energy of $Cl_2$ from –0.15 to –0.04 kcal/mol. If we switch to W4.4,[19] that is, improve the post-CCSD(T) terms, this adds 0.05 kcal/mol to the TAE, bringing theory and experiment in complete agreement.

**CONCLUSIONS**

The main conclusions we can draw from these results are the following:

- For the valence CCSD component, which represents the lion's share of the basis set convergence problem in computational thermochemistry, there are significant differences between orbital-based CCSD/AV{5,6}Z+d binding energies and their CCSD-F12b/cc-pV{Q,5}Z-F12 counterparts.

- Upon exploration of radially more flexible basis set families in the orbital CCSD calculations, these differences are greatly reduced. Even the addition of a single zeta in just the valence orbitals removes most of the discrepancy.

- The effect is particularly pronounced for second-row compounds, and to a lesser extent for first-row compounds with strongly ionic bonds.

- Counterpoise calculations reveal that, while TAEs with V5Z-F12 basis sets are nearly free of BSSE, orbital calculations have significant BSSE even with AV(6+d)Z basis sets. AV{5,6}Z+d extrapolation still leaves BSSE in the 0.10 kcal/mol range for $Cl_2$. The problem is greatly reduced by switching to ACV{5,6}Z core-valence basis sets, or the next larger valence basis sets with the top angular momentum deleted.

- In F12 calculations, the main advantage of cc-pVnZ-F12 basis sets (n=T,Q,5) over the AVnZ counterparts rests in greatly reduced basis set superposition error. BSSE in fact causes nonmonotonic basis set convergence of the atomization energy with AVnZ basis sets.

- Even AV(6+d)Z, let alone AV(5+d)Z, basis sets still do not reach the SCF limit for some second-row systems. Switching to ACV6Z or even ACV5Z completely removes the issue.

- Previous reports that all-electron approaches like HEAT lead to different CCSD(T) limits than "valence limit+CV correction" approaches like W4 theory can be rationalized in terms of the greater radial flexibility of core-valence basis sets.

- Considering the great cost and mass storage requirements of ACV6Z basis set calculations, CCSD-F12b/cc-pV{Q,5}Z-F12 offers an accurate and cost-effective alternative, as demonstrated by an application on benzene.

- For (T) corrections, however, Marchetti-Werner scaling or the CCSD-F12b based variant proposed by us[70] still cause unacceptable errors, while the term can be obtained accurately and fairly inexpensively from conventional calculations with at most AV(5+d)Z basis sets.

- At the end of the day, for the W4-F12 protocol, we recommend obtaining the SCF and valence CCSD components from CCSD-F12b/cc-pV{Q,5}Z-F12 calculations, but the (T) component from conventional CCSD(T)/aug'-cc-pV{Q,5}Z+d calculations using Schwenke's extrapolation. W4-F12 is found to yield better agreement with ATcT reference data than ordinary W4, despite W4-F12 having much smaller CPU time and resource requirements.

**Author Contributions**

The manuscript was written through contributions of all authors. All authors have given approval to the final version of the manuscript.

**ACKNOWLEDGMENTS**.

NS acknowledges a Feinberg Graduate School fellowship. Research at Weizmann was supported by the Israel Science Foundation (grant 1358/15), the Minerva Foundation, the Lise Meitner-Minerva Center for Computational Quantum Chemistry, and a grant from the Yeda-Sela Initiative (Weizmann Institute of Science). AK acknowledges an Australian Research Council (ARC) Discovery Early Career Researcher Award (DECRA, project number: DE140100311).

**Supporting Information**. V5Z-F12 and V5Z-F12(rev2) basis sets for H, B–Ne, and Al–Ar in machine-readable format. Cartesian coordinates for the 14 additional species in the W4-15 dataset.